\documentclass[twocolumn,showpacs,preprintnumbers,amsmath,amssymb,aps,prl,superscriptaddress]{revtex4-2}
\usepackage[dvipdfmx]{graphicx}
\usepackage{dcolumn}
\usepackage{bm}
\usepackage{color} 
\usepackage{ulem} 

\usepackage{xspace}

\begin{document}

\title[Anisotropic Non-Fermi Liquid and Dynamical Planckian Scaling of a Quasi-Kagome Kondo Lattice System]{Anisotropic Non-Fermi Liquid and Dynamical Planckian Scaling of a Quasi-Kagome Kondo Lattice System}

\author{Shin-ichi~Kimura}
\email{kimura.shin-ichi.fbs@osaka-u.ac.jp}
\affiliation{Graduate School of Frontier Biosciences, Osaka University, Suita, Osaka 565-0871, Japan}
\affiliation{Department of Physics, Graduate School of Science, Osaka University, Toyonaka, Osaka 560-0043, Japan}
\affiliation{Institute for Molecular Science, 
Okazaki, Aichi 444-8585, Japan}
\author{Muhammad~Frassetia~Lubis}
\affiliation{Department of Physics, Graduate School of Science, Osaka University, Toyonaka, Osaka 560-0043, Japan}
\author{Hiroshi~Watanabe}
\affiliation{Graduate School of Frontier Biosciences, Osaka University, Suita, Osaka 565-0871, Japan}
\affiliation{Department of Physics, Graduate School of Science, Osaka University, Toyonaka, Osaka 560-0043, Japan}
\author{Yasuyuki~Shimura}
\affiliation{Department of Quantum Matter, AdSE, Hiroshima University, Higashi-Hiroshima, Hiroshima, 739-8530, Japan}
\author{Toshiro~Takabatake}
\affiliation{Department of Quantum Matter, AdSE, Hiroshima University, Higashi-Hiroshima, Hiroshima, 739-8530, Japan}
\date{\today}

\begin{abstract}
At the quantum critical point of correlated materials, a non-Fermi liquid state appears where electron correlations continuously develop to very low temperatures. The relaxation time of the interacted electrons, namely quasiparticles, is scaled with the Planckian time, $\hbar/k_{\rm B}T$. However, there is a debate over whether heavy-fermion systems can obey the Planckian time. In the optical conductivity spectra, the Drude response will appear as the scaling of $\hbar\omega/k_{\rm B}T$ as the dynamical Planckian scaling (DPS). Here, we report the non-Fermi liquid behavior in the Drude response of a candidate for such materials, the quasi-kagome Kondo lattice CeRhSn. Even though the material shows a strong valence fluctuation, renormalized Drude responses observed at the photon energy below 100~meV are characterized by non-Fermi-liquid-like scattering rate $1/\tau$. The heavy carriers' Drude response only for the Ce quasi-kagome plane obeyed DPS below 80~K, suggesting the anisotropic quantum criticality with the strong $c \textrm{-} f$ hybridization.

\end{abstract}

%
\maketitle
%
\section{Introduction}
Characteristic physical properties such as non-BCS superconductivity and giant magnetoresistance emerge near the quantum critical point (QCP) of strongly correlated electron systems~\cite{Sachdev2011-tq}.
These properties originate from the many-body effect of localized and conduction electron spins, of which heavy-fermion systems and copper-oxide high-$T_c$ superconductors are typical examples.
In both materials, strongly correlated quasiparticles appear on the itinerant side of the QCP, and Landau's Fermi-liquid theory explains their behavior~\cite{Landau1959-br}.
On the other hand, magnetism appears on the localized side of the QCP due to exchange interactions between localized spins.
At this boundary, near the QCP, spin fluctuations dominate, and various properties originating from the strong correlation emerge.

Such strong electron correlation can be regarded as quantum entanglement.
Recent theoretical works have been developed at the cross-points of condensed matter, elementary-particle physics, and quantum-information theory~\cite{Hartnoll2010-wk, Zaanen2019-ra}.
These developments claim that simple principles, namely Planckian dissipation, may surprisingly govern the physics of such matter, where the relaxation time of quasiparticles of strongly correlated electron systems is determined as the Planckian time $\hbar/k_{\rm B} T$~\cite{Hartnoll2022-tq}.
This phenomenon mainly manifests in fundamental physical properties, such as a linear increase in electrical resistivity with temperature~\cite{Zaanen2019-ra,Bruin2013-mi,Grissonnanche2021-ew,Mousatov2021-xo,Guo2021-kx}.
It is also expected to appear in many physical quantities, for instance, its relationship to self-energies observed in photoelectron spectra and the Drude response in optical conductivity [$\sigma_1(\omega)$] spectra~\cite{Hartnoll2022-tq}.
The Drude response is discussed to scale (Dynamical Planckian Scaling: DPS) with the photon energy normalized by temperature ($\hbar \omega / k_{\rm B} T$)~\cite{Horowitz2012-aj,Li2023-ug}, and has been reported in high-$T_c$ cuprates~\cite{Van_der_Marel2003-ib,Michon2023-yh} and a heavy-fermion material YbRh$_2$Si$_2$~\cite{Prochaska2020-wf}.
However, it is currently debated whether quasiparticles in heavy-fermion systems are Planckian or not~\cite{Taupin2022-jw}.

One of those predicted to follow the DPS is CeRhSn~\cite{Kandala2022-rj}.
CeRhSn is a valence-fluctuation material~\cite{Gamza2009-pm,Niehaus2015-jr} with a hexagonal ZrNiAl-type crystal structure (No.~189, $P\bar{6}2m$)~\cite{Pottgen2015-bn} (shown in Fig.~\ref{fig:band}(b)).
The Ce atoms assemble a quasi-kagome lattice in the $ab$-plane.
The electrical resistivity has a large anisotropy, {\it i.e.}, the electrical resistivity along the $a$-axis is about $3-5$ times higher than that along the $c$-axis,
in contrast to the low resistivity ratio of 1.4 at most in LaRhSn without $4f$ electrons~\cite{Kim2003-az}.
The large anisotropy in CeRhSn originates from anisotropic magnetic interactions, and the material is located near the antiferromagnetic instability~\cite{Tou2004-lp}.
Non-Fermi-liquid (NFL) behavior appears below 1~K in the specific heat and thermal expansion, originating from this geometric frustration in the quasi-kagome structure of the Ce ions~\cite{Tokiwa2015-hm}.
The anisotropic spin fluctuations are also essential for the NFL behavior~\cite{Kittaka2021-vr}.
It should be investigated how the anisotropic geometrical frustration affects the electronic structure and whether the renormalized Drude response follows the DPS.

Here, we report the significantly different Drude response of the polarized $\sigma_1(\omega)$ spectra of CeRhSn along the $a$- and $c$-axes.
The results will be discussed regarding the interplay between the valence fluctuation behavior due to the $c \textrm{-} f$ hybridization and the NFL character of the quasiparticles.
Generally, $\sigma_1(\omega)$ spectra are sensitive to the $c \textrm{-} f$ hybridization and the heavy quasiparticles' behaviors~\cite{Kimura2016-nv, Kimura2021-gl, Kirchner2020-cw}.
We found that the polarized $\sigma_1(\omega)$ spectra of CeRhSn present strong anisotropies in the electronic structure and the Drude response along the $a$- and $c$-axes.
Along both axes, the $4f$ spin-orbit doublet forming $c \textrm{-} f$ hybridization appears at photon energies ($\hbar\omega$) of about 300 and 700~meV, even at room temperature, indicating strong valence fluctuation.
On the other hand, a Drude response below 100~meV reflects the formation of heavy quasiparticles. 
As the temperature decreases, the Drude weight moves to the low-energy side in both axes, suggesting a renormalization of heavy quasiparticles. 
The scattering rate $1/\tau$ derived from the extended Drude analysis is proportional to $\omega^{1}$ for $E \parallel a$ and $\omega^{1.5}$ for $E \parallel c$ at low temperatures, suggesting the NFL character.
The heavy quasiparticles' $\sigma_1(\omega)$ component obeys the DPS below 80~K only for $E \parallel a$, which is in the quasi-kagome $ab$-plane.
These results imply that the magnetic fluctuation 
of the quasi-kagome Ce lattice in the $ab$-plane
strongly couples to charge carriers via the $c \textrm{-} f$ hybridization and induces the quantum criticality.

\section{Results and Discussion}

\begin{figure*}
\centering
\begin{center}
\includegraphics[width=0.7\textwidth]{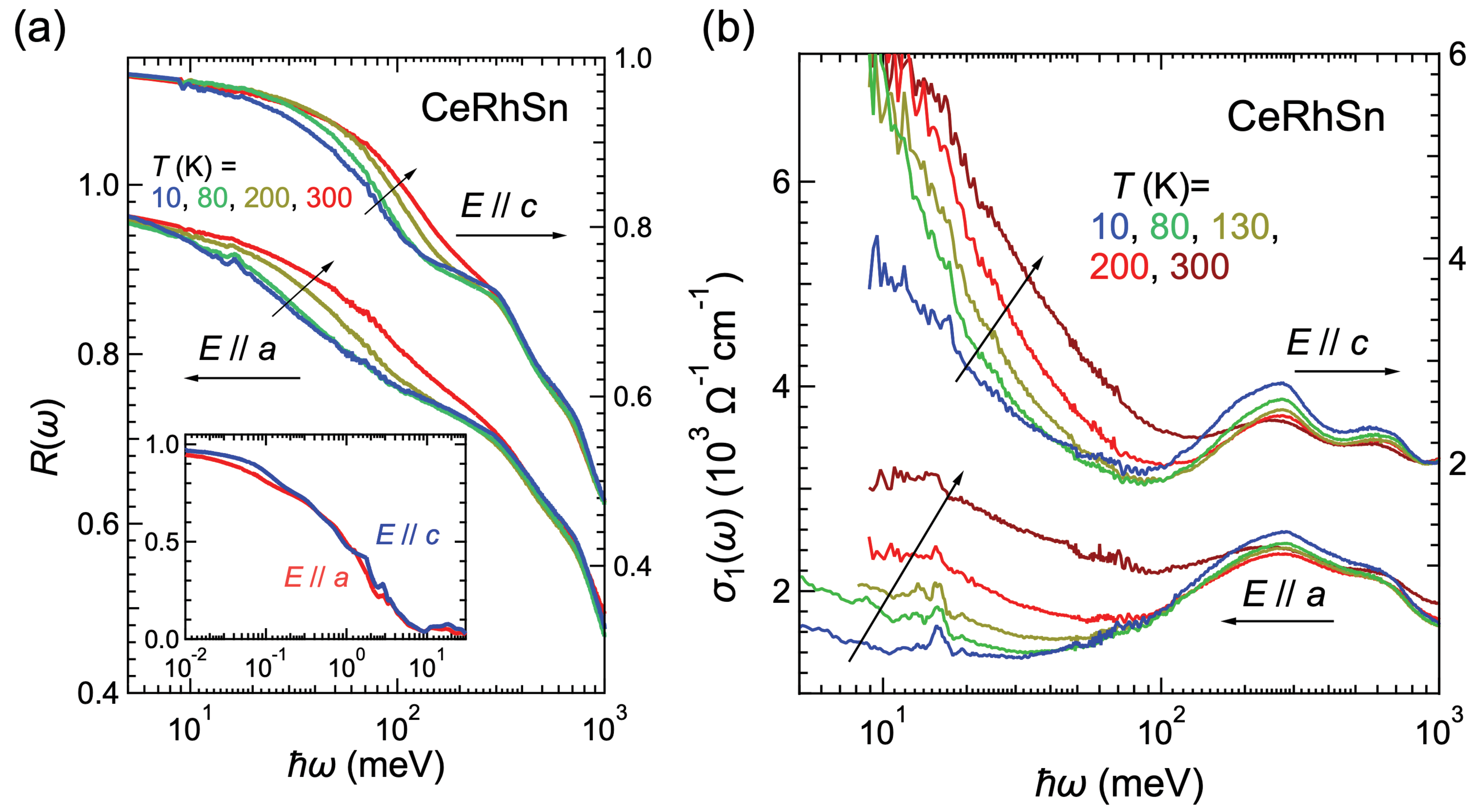}
\end{center}
\caption{
{\bf Polarized reflectivity and optical conductivity spectra of CeRhSn.}
(a) Temperature-dependent polarized reflectivity [$R(\omega)$] spectra of CeRhSn in the photon energy $\hbar\omega$ range of $5-1000$~meV.
Inset: Wide-range $R(\omega)$ spectra up to $30$~eV at 300~K.
(b) Temperature-dependent optical conductivity [$\sigma_1(\omega)$] spectra of CeRhSn with $E \parallel a$ (bottom) and $E \parallel c$ (top).
The peaks at $\hbar\omega$ $\sim15$~meV in both axes originate from phonons.
}
\label{fig:OC}
\end{figure*}

The temperature dependence of the $R(\omega)$ and $\sigma_1(\omega)$ spectra obtained from the Kramers-Kronig analysis of the $R(\omega)$ spectra of CeRhSn along the $a$- and $c$-axes is shown in Figs.~\ref{fig:OC}(a) and \ref{fig:OC}(b), respectively.
The significant axial dependence of the spectra reflects the anisotropy of the electronic state.
The $4f$ spin-orbit doublet at $\hbar\omega \sim 300$ and 700~meV, namely mid-IR peaks, originating from the strong $c \textrm{-} f$ hybridization~\cite{Kimura2016-nv}, appears in both directions, even at 300~K~\cite{Note1}.
On the other hand, a clear anisotropic Drude response appears at $\hbar\omega$ $\leq 100$~meV, which is consistent with the anisotropic electrical resistivity~\cite{Kim2003-az}.
As LaRhSn has a weak anisotropy in the electrical resistivity, the temperature dependence of $R(\omega)$ and $\sigma_1(\omega)$ spectra for $E \parallel a$ and $E \parallel c$ are also very similar as shown in Fig.~S1(a) (See Supplementary Information).
This suggests that the axial dependence of this Drude structure is attributed to the anisotropic magnetic interactions.


\subsection{Valence fluctuation observed with mid-IR peaks}

\begin{figure*}
\begin{center}
\includegraphics[width=1.0\textwidth]{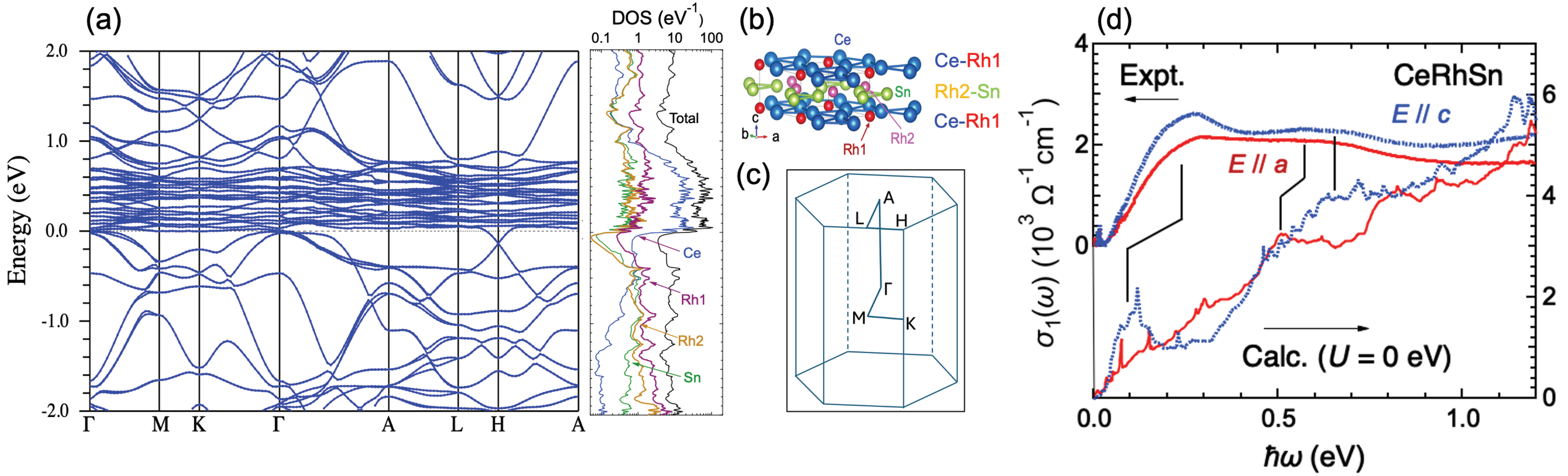}
\end{center}
\caption{
{\bf Mid-IR peaks compared to the band calculations.}
(a) Band structure and density of states (DOS) of CeRhSn by the LDA calculation with spin-orbit interaction.
(b) Crystal structure of CeRhSn with a quasi-kagome Ce lattice in the basal plane.
Rh atoms have two different sites, namely Rh1 and Rh2, corresponding to the different densities of states shown in (a).
(c) The first Brillouin zone and high symmetry points of CeRhSn.
(d) Calculated $\sigma_1(\omega)$ spectra for $E \parallel a$ (red solid line) and $E \parallel c$ (blue dashed line) compared with the experimentally obtained $\sigma_1(\omega)$ spectra after subtraction of the Drude component shown in Fig.~S2 in the supplementary materials.
}
\label{fig:band}
\end{figure*}

Firstly, the mid-IR peaks are compared to the band structure calculations.
Figure~\ref{fig:band}(a) shows the band structure along high symmetry points of CeRhSn shown in Fig.~\ref{fig:band}(c).
In the right, the density of states (DOS) by the LDA calculations is presented.
The high DOS in the range of 0~eV (= Fermi energy; $E_{\rm F}$) -- 0.6~eV originates from the Ce~$4f$ unoccupied states.
The calculated band structure is regarded as fully itinerant; therefore, the itinerant character can be checked by comparing the experimental $\sigma_1(\omega)$ spectra to the band calculation~\cite{Kimura2021-gl}.
In Figure~\ref{fig:band}(d), the $\sigma_1(\omega)$ spectra at 10~K with mid-IR peaks at about 0.2--0.3 and 0.6--0.7~eV are compared to the calculated $\sigma_1(\omega)$ spectra.
Significant peaks at $\sim0.1$ and $\sim0.5$~eV correspond to the experimentally observed mid-IR peaks, although the energy is shifted. 
A similar shift in energy was reported in the itinerant superconductor CeRh$_2$As$_2$~\cite{Kimura2021-dn}.
Therefore, in addition to the appearance of the mid-IR peaks at 300~K, the good correspondence of the mid-IR peaks to the calculated $\sigma_1(\omega)$ spectra confirms the strong $c \textrm{-} f$ hybridization strength~\cite{Okamura2007-si, Kimura2016-nv}, which is consistent with the results of previous photoelectron experiments~\cite{Shimada2006-ek, Sundermann2021-oe}.
It should be noted that recent DFT+DMFT calculations on CeRhSn more reproduce the mid-IR peaks~\cite{Bohm2024-lb}, which also suggests the strong $c \textrm{-} f$ hybridization intensity.


\subsection{Extended Drude analysis}

\begin{figure*}
\begin{center}
\includegraphics[width=0.9\textwidth]{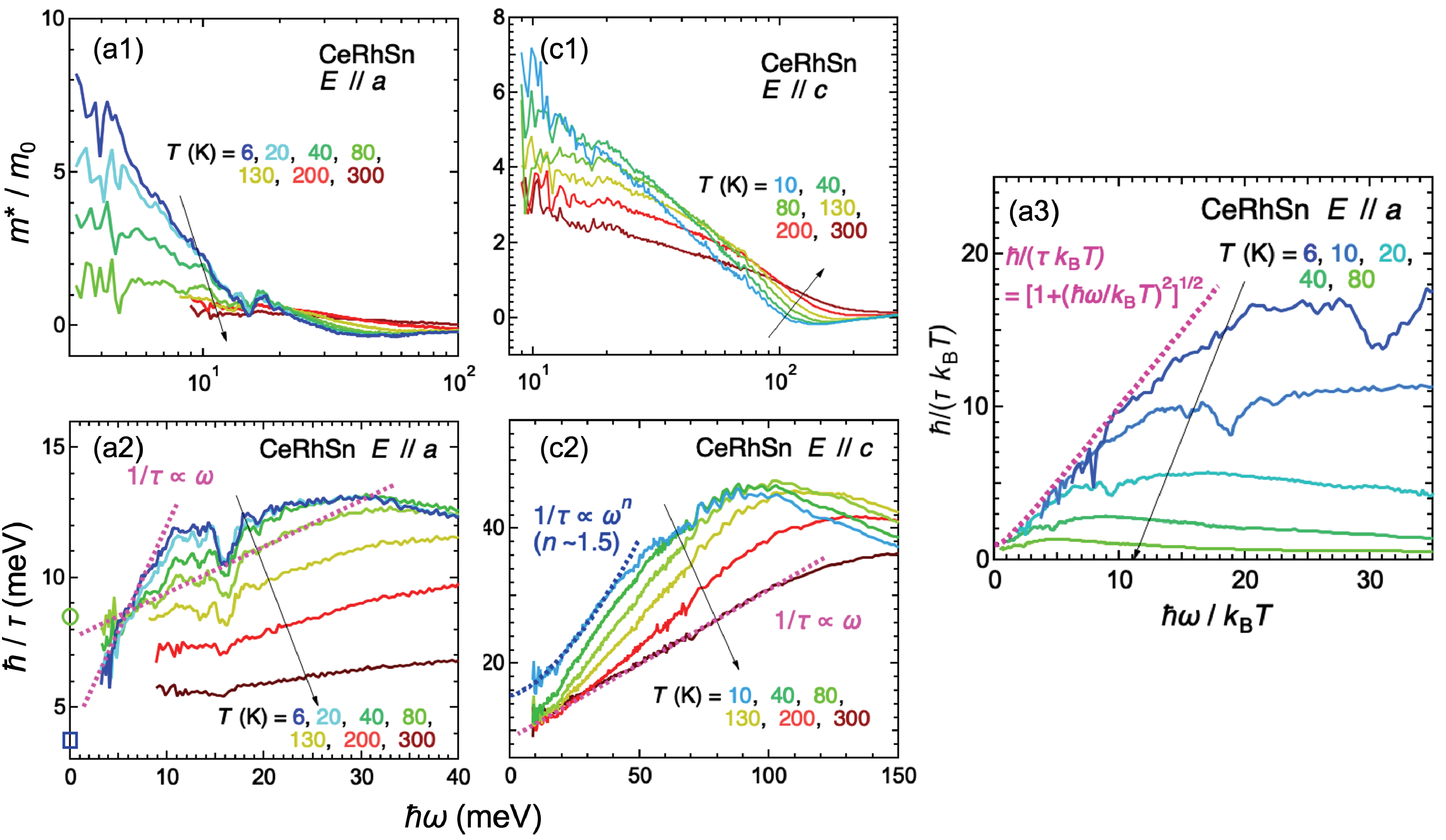}
\end{center}
\caption{
{\bf Extended Drude analysis.}
(a1, a2) Mass enhancement ($m^*/m_0$, a1) and scattering rate ($\hbar/\tau$, a2) as functions of photon energy and temperature for $E \parallel a$.
$\hbar/\tau$ values at $\hbar\omega = 0~{\rm eV}$ evaluated by the electrical resistivity and Hall coefficient data~\cite{Kim2003-az} are plotted with an open circle (80~K) and an open square (6~K) in (a2).
(a3) 
$\hbar/(\tau k_{\rm B}T)$ as a function of $\hbar\omega/k_{\rm B}T$ obtained from (a2), but the residual damping of $\hbar/\tau = 3.3~{\rm meV}$ obtained from the electrical resistivity and the Hall coefficient is subtracted.
The formula expected from the Planckian dissipation, $\hbar/(\tau k_{\rm B}T) = [1+(\hbar\omega/k_{\rm B}T)^2]^{1/2}$, is also plotted as a dashed line.
(c1, c2) Same as a1 and a2, but for $E \parallel c$.
Note that the negative values and approaching zero of $m^*/m_0$ in (a1) and (c1) have no meaning because they appear in the photon energy regions outside of Drude components as shown in Fig.~1(b).
}
\label{fig:ExtendedDrude}
\end{figure*}

Next, we discuss the spectral shape of the Drude component using the extended Drude analysis~\cite{Kimura2006-pd}.
Figure~\ref{fig:ExtendedDrude} indicates the obtained mass enhancement ($m^*/m_0$, where $m^*$ and $m_0$ are the effective mass of the quasiparticles and the rest mass of an electron, respectively) and the scattering rate ($1/\tau$) along both directions.
Here, the carrier densities along both axes for the extended Drude analysis were evaluated using the Hall coefficients at the lowest accessible temperature of 6~K~\cite{Kim2003-az}, {\it i.e.}, $6.9\times10^{20}~{\rm cm}^{-3}$ for $E \parallel a$ and $4.2\times10^{21}~{\rm cm}^{-3}$ for $E \parallel c$.
In the low-energy limit, the effective mass increased with decreasing temperature in both directions.
However, $m^*/m_0$ for $E \parallel c$ increased continuously on cooling from 300~K, whereas $m^*/m_0$ for $E \parallel a$ is almost unchanged down to 80~K, below which $m^*/m_0$  increased significantly.
Also, in $1/\tau$, the peak energy at the lowest temperature is about 30~meV for $E \parallel a$ and about 100~meV for $E \parallel c$.
The peak energy corresponds to the energy at which $m^*/m_0$ begins to increase.
These facts suggest the different characteristics of quasiparticles depending on the crystal axis.
At the lowest temperature, $1/\tau$ is proportional to $\omega^1$ for $E \parallel a$.
On the other hand, for $E \parallel c$, $1/\tau$ is roughly proportional to $\omega^{1.5}$, which is not $\omega^1$ nor $\omega^2$.
It is known that $1/\tau$ is proportional to $\omega^2$ in the normal Fermi liquid state~\cite{Kimura2006-dc}, but it is proportional to $\omega^1$ if the state is located very near QCP~\cite{Kimura2006-pd}.
In the case that the state is slightly shifted from QCP but in an NFL state, $1/\tau$ is proportional to $\omega^n$ with $1<n<2$~\cite{Iizuka2010-ov}.
Therefore, the power dependence of $1/\tau$ on $\omega$ indicates that the state along the $a$ axis is located very near QCP and is slightly shifted from QCP along the $c$ axis, but both axes are in NFL states.
This result is consistent with the recently observed anisotropic NFL behavior in the specific heat~\cite{Kittaka2021-vr}.
In CeRhSn, the $T$-linear region is below 40~K ($\sim~3~{\rm meV}$).
Even at 80~K, $1/\tau$ is proportional to $\hbar\omega$ as shown in Fig.~\ref{fig:ExtendedDrude}(a2).
This fact suggests that the Planckian form can be applied at temperatures below 80~K.

In Planckian metals of NFL, $\hbar/\tau$ will follow $[(\hbar \omega)^2+(k_{\rm B} T)^2]^{1/2}$, i.e., $\hbar/(\tau k_{\rm B}T) \sim [1+(\hbar\omega/k_{\rm B}T)]^{1/2}$.
The experimental $\hbar/(\tau k_{\rm B}T)$ subtracted by the residual damping $\hbar/\tau(0) = 3.3~{\rm meV}$ evaluated from the extrapolated values of the electrical resistivity and Hall coefficient to 0~eV~\cite{Kim2003-az} are plotted as a function of $\hbar\omega/k_{\rm B}T$ in Fig.~\ref{fig:ExtendedDrude}(a3).
The slope in the region of $\hbar\omega/k_{\rm B}T \leq 10$ is scaled with $\hbar\omega/k_{\rm B}T$.
The ideal Planckian formula $\hbar/(\tau k_{\rm B}T) = [1+(\hbar\omega/k_{\rm B}T)^2]^{1/2}$ is also plotted in the figure.
The formula can explain the slope, suggesting that the $E \parallel a$ of CeRhSn is Planckian.

For $E \parallel a$, as shown in Fig.~\ref{fig:ExtendedDrude}(a2), the $\hbar/\tau$ at $\hbar\omega = 0~{\rm eV}$ decreases with decreasing temperature, which is the definition of Planckian metals, but that at $\hbar\omega \sim 10~{\rm meV}$ shows the opposite behavior, i.e., the slope ($d \tau^{-1} / d\omega$) is strongly suppressed with increasing temperature.
This result is in contrast to the parallel slope of $\hbar/\tau(\omega)$ at different temperatures in a high-$T_c$ cuprate~\cite{Michon2023-yh}.
However, as shown in Fig.~\ref{fig:ExtendedDrude}(a3), the constant slope of $d(\tau^{-1}T^{-1})/d(\omega/T)$ at different temperatures in the region of $\hbar\omega/k_{\rm B}T \leq 10$ following the Planckian scaling might be a property of heavy fermion materials.


\subsection{Dynamical Planckian scaling}

\begin{figure*}
\begin{center}
\includegraphics[width=0.7\textwidth]{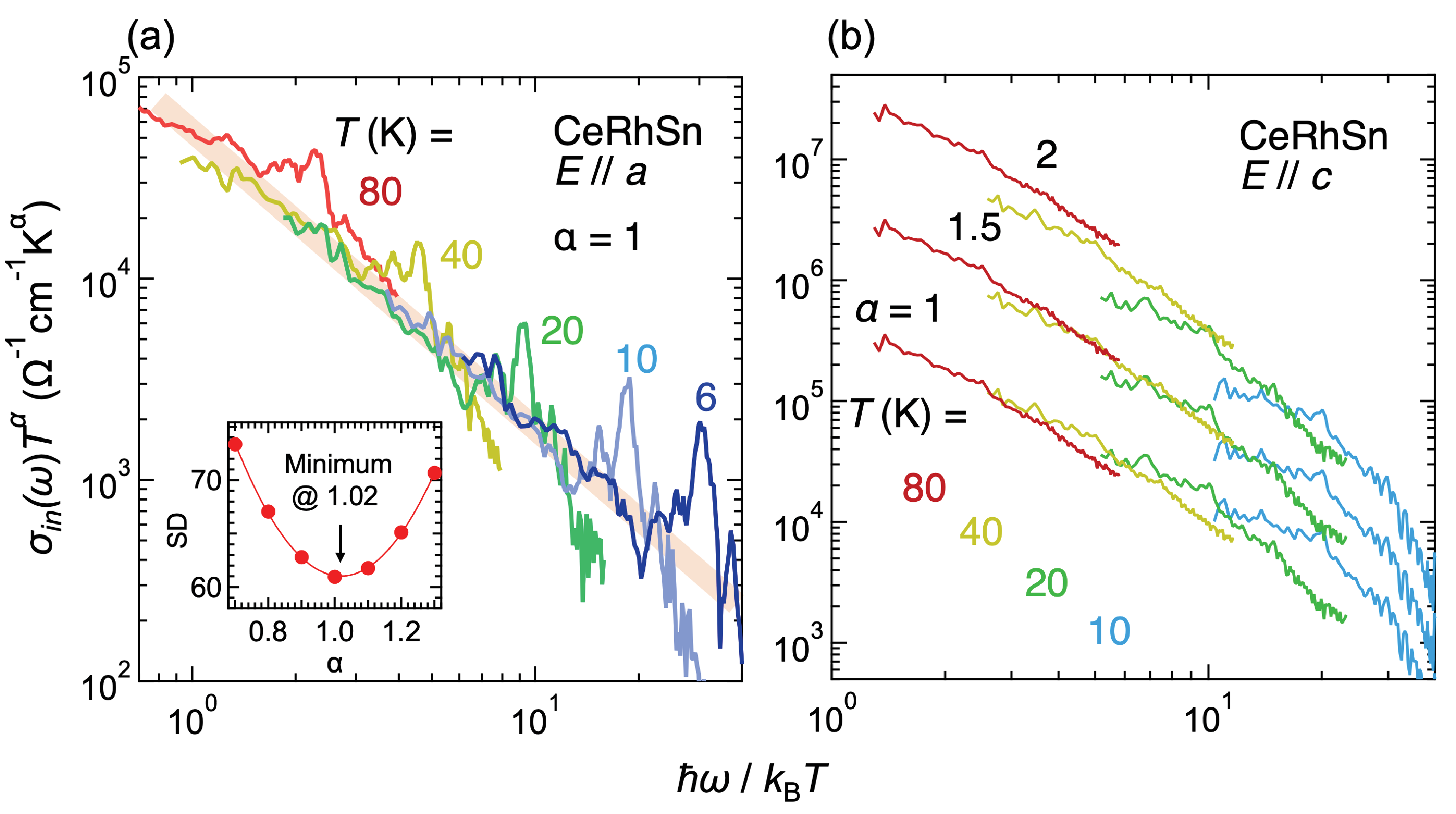}
\end{center}
\caption{
{\bf Dynamical Planckian scaling plot.}
$\sigma_{in}(\omega) \cdot T^\alpha$
as a function of $\hbar \omega / k_{\rm B} T$  as shown in Eq.~\ref{eq:DPS} for $E \parallel a$ (a) and $E \parallel c$ (b).
The intrinsic heavy quasiparticles' $\sigma_{in}(\omega)$ spectra were obtained by subtracting the background spectra due to light quasiparticles and interband transitions from the original $\sigma_1(\omega)$ spectra.
The broad solid line in (a) indicates a guide for eyes to indicate the scaling.
DPS requires $\alpha=1$, but different $\alpha$ values are adopted for $E \parallel c$ in (b).
The inset of (a) shows the standard deviation (SD) of $\log[\sigma_{in}(\omega)T^\alpha]$ vs $\log[\hbar\omega/k_{\rm B}T]$ from a straight line as a function of $\alpha$.
}
\label{fig:DPS}
\end{figure*}

Finally, we discuss whether the temperature dependence of the Drude component can be explained with DPS.
If DPS is realized, the temperature dependence of the $\sigma_1(\omega)$ spectrum has a relationship~\cite{Li2023-ug,Prochaska2020-wf}:

\begin{equation}
\sigma_{in}(\omega) \cdot T^\alpha = f(\hbar \omega/k_{\rm B}T) , 
\label{eq:DPS}
\end{equation}
where $\sigma_{in}(\omega)$ is the real-part intrinsic optical conductivity and $f(x)$ is a function and $\alpha=1$ for DPS.

Here, the $\sigma_1(\omega)$ spectra in Fig.~\ref{fig:OC}(b) comprise carriers' contributions and interband transitions.
The carrier component can be classified as ``heavy'' and ``light'' quasiparticles with and without strong electron correlation.
On the other hand, the second contribution originates from the interband transition from the valence band to the $c \textrm{-} f$ hybridization band with the Ce~$4f$ spin-orbit splitting.
Under these assumptions, the $\sigma_1(\omega)$ spectrum is decomposed into two Drude ($\sigma_{in}(\omega)$ and $\sigma_{BG}(\omega)$ for the heavy and light quasiparticle components, respectively) and two Lorentz (corresponding to the Ce~$4f_{5/2}$ amd $4f_{7/2}$ states) components as shown in Fig.~S2.
The individual contributions from heavy and light quasiparticles manifest in the several effective masses appearing in quantum oscillations~\cite{Onuki2004-cy}.
In $\sigma_1(\omega)$ spectra, the evidence of the light quasiparticles appears as a background of the Drude structure~\cite{Awasthi1993-yv,Degiorgi1999-rj}.
Then, we derive the heavy quasiparticles' intrinsic optical conductivity $\sigma_{in}(\omega)$ spectra from the $\sigma_1(\omega)$ spectra by subtracting the light quasiparticles' background $\sigma_{BG}(\omega)$ and two Lorentzians.
The DPS plot of $\sigma_{in}(\omega)$ is shown in Fig.~\ref{fig:DPS}(a, b).
(The DPS plot for all temperatures for $E \parallel a$ is shown in Fig.~S3 in the Supplementary Information.)
In $E \parallel a$, the standard deviation (SD) from a straight line to $\log \sigma_{in}(\omega)T^\alpha$ vs $\log \hbar\omega/k_{\rm B} T$ at different $\alpha$ values is plotted in the inset of Fig.~\ref{fig:DPS}.
The minimum SD appears at $\alpha \sim 1.02$, which is nearly equal to $1$.
The result is consistent with the case of YbRh$_2$Si$_2$~\cite{Prochaska2020-wf}, suggesting a universal behavior in heavy fermion materials.
The DPS is realized at temperatures below 80~K for $E \parallel a$.
However, for $E \parallel c$, the data are not scaled even with $\alpha = 1.5$ and $2$.
Therefore, it could be concluded that the heavy quasiparticles for $E \parallel a$ follow the DPS.


The maximum electrical resistivity along the $a$-axis appears at about 80~K, below which the coherent Kondo lattice is realized~\cite{Kim2003-az}.
The fact suggests that the DPS for $E \parallel a$ appears in the coherence state. 
It would be due to the geometrical frustration of the quasi-kagome structure.
However, in the temperature regions where the DPS holds in Fig.~\ref{fig:DPS}(a), the relation between $\log[\sigma_{in}(\omega) \cdot T]$ and $\log[\hbar \omega/k_{\rm B}T]$ is a straight line with the relation of $\sigma_{in}(\omega) \cdot T \propto (\hbar \omega/k_{\rm B}T)^{-1.8}$.
In high-$T_c$ cuprates, $\sigma_1(\omega, T)\cdot T$ is a universal function of $\omega/T$ based on the Drude formula~\cite{Van_der_Marel2003-ib}.
The formula for $\hbar \omega /k_{\rm B}T \gg 1$ becomes $\sigma_1(\omega, T)\cdot T \propto (\hbar \omega/k_{\rm B}T)^{-2}$.
The relation of $\sigma_{in}(\omega) \cdot T \propto (\hbar \omega/k_{\rm B}T)^{-1.8}$ for CeRhSn is almost consistent with the simple formula.
However, the order is inconsistent with the value of YbRh$_2$Si$_2$, where $\sigma_{in}(\omega)T \propto (\hbar \omega/k_{\rm B}T)^{-1}$~\cite{Prochaska2020-wf}.
These results imply that the heavy fermion systems of CeRhSn and YbRh$_2$Si$_2$ follow the DPS but have individual scaling, which may provide a meaningful result.
Further data should be accumulated to establish a universal DPS in heavy-fermion systems.

As shown in Fig.~\ref{fig:ExtendedDrude}(a3), the $\hbar/(\tau k_{\rm B} T)$ curve can follow the Planckian scaling of $\hbar/(\tau k_{\rm B}T) \sim [1+(\hbar\omega/k_{\rm B}T)]^{1/2}$ and $\hbar/(\tau k_{\rm B} T) \sim 1$ at $\hbar \omega = 0$, which is consistent with quantum oscillation data, where $\hbar/(\tau k_{\rm B} T) \sim 1$ is observed~\cite{Bruin2013-mi}.
The same carriers can probably be observed in quantum oscillations and optical conductivity experiments. 
On the other hand, thermodynamical properties at $E_{\rm F}$ pronounce $\hbar/(\tau k_{\rm B} T) = 0.01 - 0.02$, where the Planckian dissipation is under debate~\cite{Taupin2022-jw}.
The inconsistency in $\hbar/(\tau k_{\rm B} T)$ should be resolved by the results of many further experiments.

In most three-dimensional Ce-based heavy-fermion materials in the vicinity of QCP, the mid-IR peaks are slightly visible owing to the relatively weak $c \textrm{-} f$ hybridization intensity~\cite{Marabelli1990-pf, Singley2002-ke, Kimura2009-tv, Kimura2016-nv}.
However, the mid-IR peaks clearly appear in CeRhSn, being the hallmark of strong valence fluctuation.
The simultaneous appearance of the valence fluctuation and DPS owing to the quasi-kagome structure describes that the quantum criticality of CeRhSn is different from usual NFL heavy-fermions like CeCu$_{6-x}$Au$_x$, where AFM correlations are responsible for the quantum criticality~\cite{Lohneysen2007-bi}.

%
\subsection{Conclusion remarks}

To summerize, polarized optical conductivity measurements and first-principles calculations of the quasi-kagome Kondo lattice material CeRhSn have revealed the anisotropic electronic structure and Drude response.
Along the hexagonal $a$- and $c$-axes, the $4f$ spin-orbit doublet showing the strong $c \textrm{-} f$ hybridization appears even at room temperature, indicating the strong $c \textrm{-} f$ hybridization intensity. 
On the other hand, a renormalized Drude response at $\hbar\omega$~$\leq 100$~meV indicates the formation of heavy quasiparticles. 
Analysis of the Drude structure shows that it follows the DPS only along the $a$-axis, resulting from the magnetic fluctuations based on the Ce quasi-kagome lattice.
These findings support that the quantum criticality of CeRhSn coexists with the valence fluctuation.
This work should motivate further investigation to clarify whether the anisotropic DPS commonly describes low-temperature responses of low-dimensional NFL heavy-fermion materials.

\section*{Methods}
\subsection*{Sample preparation}
Single crystals of CeRhSn and a nonmagnetic counterpart LaRhSn were grown by the Czochralsky method in a radio-frequency induction furnace~\cite{Kim2003-az}.
The samples were polished to mirror surfaces with 3M$^{\rm TM}$ Lapping Film Sheets (0.3 Micron Grade) along the crystal axes to measure near-normal-incident polarized optical reflectivity [$R(\omega)$] spectra.
\subsection*{Optical conductivity measurements}
The $R(\omega)$ spectra were acquired in a wide $\hbar\omega$ range of 5~meV -- 30~eV to ensure accurate Kramers-Kronig analysis (KKA)~\cite{Kimura2013-rg}.
Infrared and terahertz measurements at $\hbar\omega$ = 5--30~meV and 0.01--1.5~eV have been performed using $R(\omega)$ measurement setups with an automatic sample positioning system at varying temperatures of 6--300~K~\cite{Kimura2008-tg}.
The absolute values of $R(\omega)$ spectra were determined with the {\it in-situ} gold evaporation method.
In the $\hbar\omega$ range of 1.5--30~eV, the $R(\omega)$ spectrum was acquired only at 300~K by using the synchrotron radiation setup at the beamline 3B~\cite{Fukui2014-wz} of UVSOR-III Synchrotron~\cite{Ota2022-ak} and connected to the spectra for $\hbar\omega$ $\leq 1.5$~eV for conducting KKA.
In order to obtain $\sigma_1(\omega)$ via KKA of $R(\omega)$, the spectra were extrapolated below 5~meV with a Hagen-Rubens function [$R(\omega)=1-\left(2\omega/(\pi \sigma_{DC})\right)^{1/2}$]
due to the metallic $R(\omega)$ spectra, and above 30~eV with a free-electron approximation [$R(\omega) \propto \omega^{-4}$]~\cite{Dressel2002-ae}.
The background dielectric constant at $\hbar\omega = 30~{\rm eV}$ was set as unity because the energy is higher than those of the valence band and the most shallow core levels of Ce~$5p$ and $5s$.
Here, the direct current conductivity ($\sigma_{DC}$) values were adopted from the experimental values~\cite{Kim2003-az}.
The extrapolations were confirmed not to severely affect the $\sigma_1(\omega)$ spectra at $\hbar\omega = 3-100~{\rm meV}$, which is the main part of this paper.
$R(\omega)$ and $\sigma_1(\omega)$ spectra of LaRhSn have been measured as a reference without $4f$ electrons, and it is shown in Fig.~S1 of Supplementary Information.
\subsection*{First-principle band calculations}
First-principle local-density approximation (LDA) calculations of the band structure have been performed by using the {\sc Wien2k} code, including spin-orbit interaction~\cite{Blaha2020-vn}.
Thereby, lattice parameters reported in Ref.~\cite{Kim2003-az} were used.
The calculated band structure (shown in Fig.~\ref{fig:band}(a)) is consistent with the previous reports~\cite{Slebarski2004-bg,Al_Alam2008-jn,Gamza2009-pm}.
The theoretical $\sigma_1(\omega)$ curves in Fig.~2(d) were obtained with the {\sc Wien2k} code based on the derivation of the dielectric tensor within the random-phase approximation~\cite{Ambrosch-Draxl2006-zf}.
The electron correlation effect was also evaluated with the LDA+$U$ calculations as shown in Fig.~S1 in Supplementary Information.
The results indicate that the LDA+$U$ calculation cannot explain the experimental $\sigma_1(\omega)$ spectra of CeRhSn.

\section*{Data availability}
The datasets generated during and/or analyzed during the current study are available from the corresponding author upon request.
\\
%
%

%
%

\bibliography{CeRhSn_Notes}

\section*{Acknowledgments}
We thank UVSOR Synchrotron staff members for their support during synchrotron radiation experiments.
Part of this work was performed under the Use-of-UVSOR Synchrotron Facility Program (Proposals No.~23IMS6016) of the Institute for Molecular Science, National Institutes of Natural Sciences.
This work was partly supported by JSPS KAKENHI (Grant Nos.~23H00090, 22K03529, 17K05545).

\section*{Author Contributions}
Y.~S. and T.~T. fabricated single crystals of CeRhSn and LaRhSn and characterized them.
S.~K., M.~F.~L., and H.~W. measured reflectivity spectra in a wide energy range from THz to VUV.
S.~K. analyzed the obtained data, performed the band structure calculations, conceived the project, and was responsible for its overall execution.
All authors discussed the results and commented on the manuscript.

\section*{Competing Interests}
The authors declare no competing interest.

\section*{Additional Information}
Supplementary information is available for this paper at https://doi.org/10.1038/s41535-025-00797-w.

\end{document}


\title{
Supplementary Information for \\
``Anisotropic Non-Fermi Liquid and Dynamical Planckian Scaling of the Quasi-Kagome Kondo Lattice CeRhSn''
}
\author{Shin-ichi~Kimura}
\email{sk@kimura-lab.com}
\affiliation{Graduate School of Frontier Biosciences, Osaka University, Suita 565-0871, Japan}
\affiliation{Department of Physics, Graduate School of Science, Osaka University, Toyonaka 560-0043, Japan}
\affiliation{Institute for Molecular Science, 
Okazaki 444-8585, Japan}
\author{Muhammad~Frassetia~Lubis}
\affiliation{Department of Physics, Graduate School of Science, Osaka University, Toyonaka 560-0043, Japan}
\author{Hiroshi~Watanabe}
\affiliation{Graduate School of Frontier Biosciences, Osaka University, Suita 565-0871, Japan}
\affiliation{Department of Physics, Graduate School of Science, Osaka University, Toyonaka 560-0043, Japan}
\author{Yasuyuki~Shimura}
\affiliation{Department of Quantum Matter, AdSE, Hiroshima University, Higashi-Hiroshima 739-8530, Japan}
\author{Toshiro~Takabatake}
\affiliation{Department of Quantum Matter, AdSE, Hiroshima University, Higashi-Hiroshima 739-8530, Japan}
\date{\today}
%
%
\maketitle
%
\section{S1.~Mid-Infrared spectra of LaRhSn and CeRhSn compared with band calculations}
%
\begin{figure*}
\includegraphics[width=0.9\textwidth]{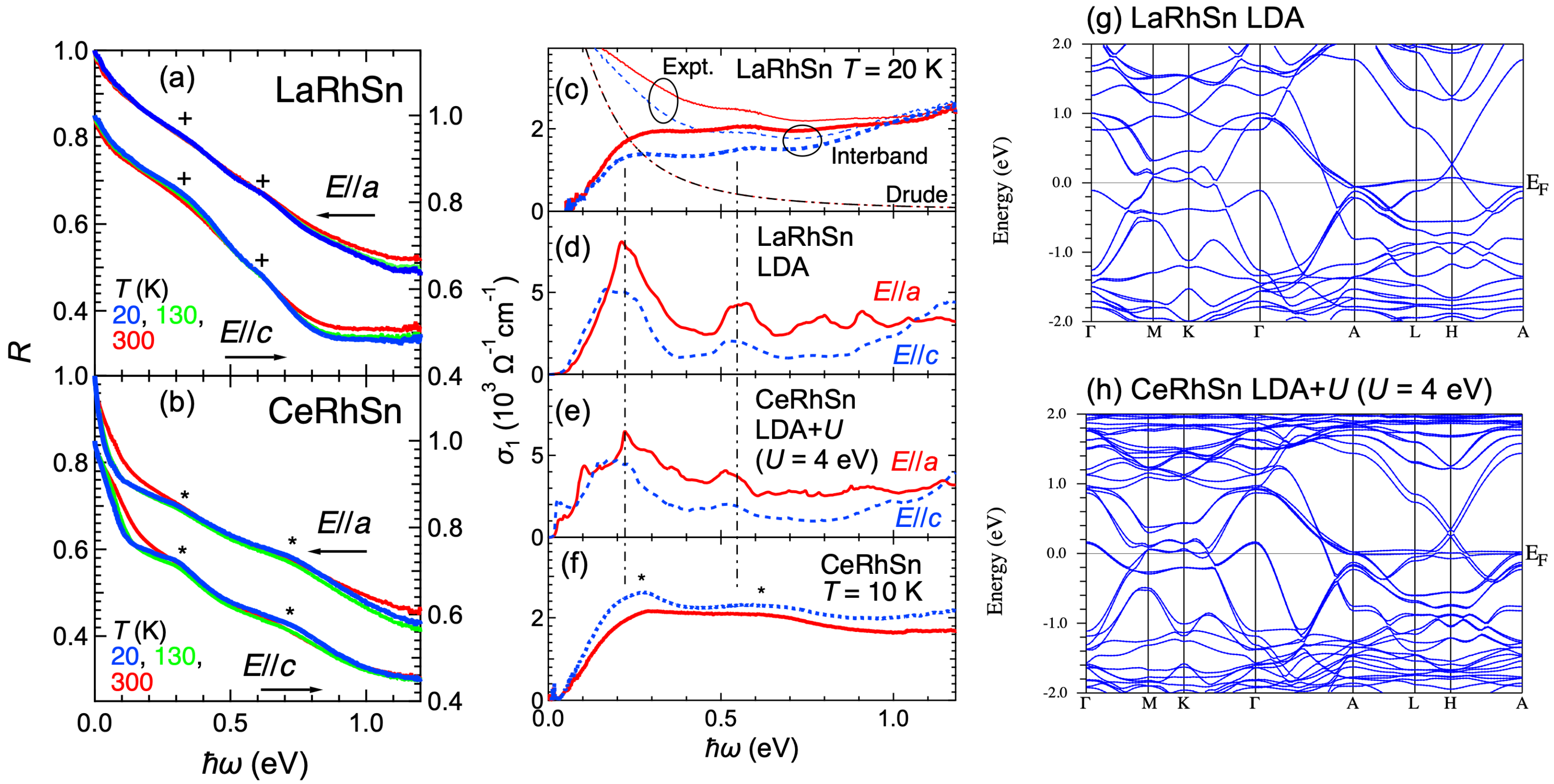}
\caption{
(a, b) Temperature-dependent reflectivity spectra $R(\omega)$ of LaRhSn and CeRhSn along two independent axes. 
(c) Optical conductivity spectra $\sigma_1(\omega)$ and the interband transition and Drude components of LaRhSn. 
(d) Theoretical anisotropic $\sigma_1(\omega)$ spectra of the interband part of LaRhSn. 
(e) Same as (d) but for CeRhSn where the localized Ce~$4f$ states are expected in the LDA+$U$ calculation. 
(f) Same as (c) but for CeRhSn and only the interband part. The spectra are the same as shown in Fig.~2(d). 
(g) Band structure of LaRhSn calculated using the local-density approximation (LDA). 
(h) Band structure of CeRhSn by the LDA+$U$ ($U$ = 4~eV) calculation.
The vertical dot-dashed lines in the figure are a guide for the eye that shows the correspondence between the experimental and theoretical peaks.
}
\label{fig:S1}
\end{figure*}

To investigate the renormalization effect of the Drude component and the appearance of the mid-IR peaks observed in the optical spectra of \CeRhSn, the optical reflectivity spectra \R of \LaRhSn and \CeRhSn are shown in Fig.~S1(a, b).
In \LaRhSn, the spectra at \hw$\sim0$~eV show almost no temperature dependence compared to those of \CeRhSn.
This is consistent with a monotonical decrease of electrical resistivity with decreasing temperature.
Two shoulder structures appear in both \LaRhSn (marked by +) and \CeRhSn (marked by *).
The intensity for \LaRhSn is smaller than that for \CeRhSn, and the energy is slightly different.
The origin of this structure is discussed using the results of band calculations.

Figure~S1(c, f) shows the anisotropic \OC spectra at low temperatures obtained from the Kramers-Kronig analysis of the \R spectra in Fig.~S1(a, b).
Figure~S1(c) shows the \OC spectra obtained from the \R spectra and the interband transition components obtained by subtracting the Drude component from the original \OC spectra.
In Fig.~S1(f), only the interband transition components of \CeRhSn are plotted, the same as the experimental spectra of Fig.~2(d).
Figure~S1(g,h) shows the band structure of \LaRhSn by the local-density approximation (LDA) and that of \CeRhSn by the LDA+$U$ ($U=4$~eV) calculation.
In the LDA+$U$ calculation, the Ce~$4f$ states are assumed as localized states, then Ce~$4f$ does not appear near \EF, but is separated at the upper and lower ends of the figure.
This band diagram differs significantly from the LDA calculation (Fig.~2(a)) in which the Ce~$4f$ bands are located very near \EF.
Since the band dispersions near \EF of \LaRhSn in Fig.~S1(g) and \CeRhSn in Fig.~S1(h) are very similar, the \OC spectra obtained from these band calculations (Fig.~S1(d,e)) are almost identical.

The calculated \OC spectra (Fig.~S1(d,e)) are compared with the experimental results (Fig.~S1(c,f)).
In the experimental \OC spectra of \LaRhSn, two peaks appear at about 0.22 and 0.55~eV in both $a$- and $c$-axis, and the peak intensity for \EparaA is larger than that for \EparaC.
Furthermore, the intensities for \EparaC and \EparaA are reversed at \hw$\sim 1$~eV.
The LDA calculation reproduces this peak position, intensity anisotropy, and intensity inversion.
This suggests that the double peak structure of \LaRhSn is derived from the band structure shown in Fig.~S1(g).

On the other hand, for \CeRhSn (in Fig. S1(f)), peak structures appear at about 0.3~eV and 0.6~eV, which are slightly shifted from the peak positions predicted by the LDA+$U$ calculation, and the experimental intensity for \EparaC is larger than that for \EparaA.
These results are inconsistent with the predictions from the LDA+$U$ calculation, which assumes the localized Ce~$4f$ character.
Even though the peak energies in the calculated \OC spectra using the LDA calculation in Fig.~2(d) do not agree with those of the experimental \OC spectra, the intensity ratio between \EparaA and \EparaC is consistent.
Therefore, the LDA calculation is more plausible for explaining the \OC spectra of \CeRhSn.
For a deep understanding of the mid-IR peak, a LDA+DMFT calculation might be suitable as described by B\"ohm {\it et al.}~\cite{Bohm2024-lb}.

\section*{S2.~Drude--Lorentz fitting of optical conductivity spectra of CeRhSn}

Figure~S2 represents the low-energy \OC spectra along two independent axes at 80 and 10~K.
The spectra consist of the Drude components due to carriers and the Lorentz components of the interband transitions.
Owing to the complex band structure, there are several kinds of carrier characters.
Then, the carrier characters are separated into two parts: heavy carriers due to the strong electron correlation and light carriers of weak correlation.
As explained above, the interband transition originates from the transition from the valence band to a \cf hybridization band because of the itinerant Ce~$4f$ character.
The \cf hybridization band splits into two bands due to the Ce~4f spin-orbit pair.
According to these assumptions, the \OC spectra can be fitted with the combination of two Drude and two Lorentz functions~\cite{Dressel2002-ae} as follows:
\begin{eqnarray}
\label{eq:DL}
\tilde\sigma(\omega)&=&\tilde\sigma_{in}(\omega) + \tilde\sigma_{BG}(\omega)
+ \sum_{j=1}^2 \tilde\sigma_{L, j}(\omega) \nonumber \\
&=& i \dfrac{N_{in} e^2}{m_0} \dfrac{1}{\omega+i/\tau_{in}}
+ i \dfrac{N_{BG} e^2}{m_0} \dfrac{1}{\omega+i/\tau_{BG}} \nonumber \\
& & + \sum_{j=1}^2 \dfrac{N_j e^2 \omega}{im_0} \dfrac{1}{(\omega_j^2-\omega^2)-i\omega/\tau_j} .
\end{eqnarray}
Here, $in$ and $BG$ indicate the ``intrinsic'' heavy carriers and ``background'' light carriers.
$N$ and $\tau$ are the electron number and the relaxation time, and $e$ and $m_0$ are the charge and rest mass of an electron. 
The experimental \OC spectra combined with the direct current conductivities from the electrical resistivity~\cite{Kim2003-az} were fitted using this formula by a nonlinear least-square fitting method with the IGOR Pro software, as shown in Fig.~S2 and the obtained parameters are shown in Table S1.

\begin{figure}
\begin{center}
\includegraphics[width=0.5\textwidth]{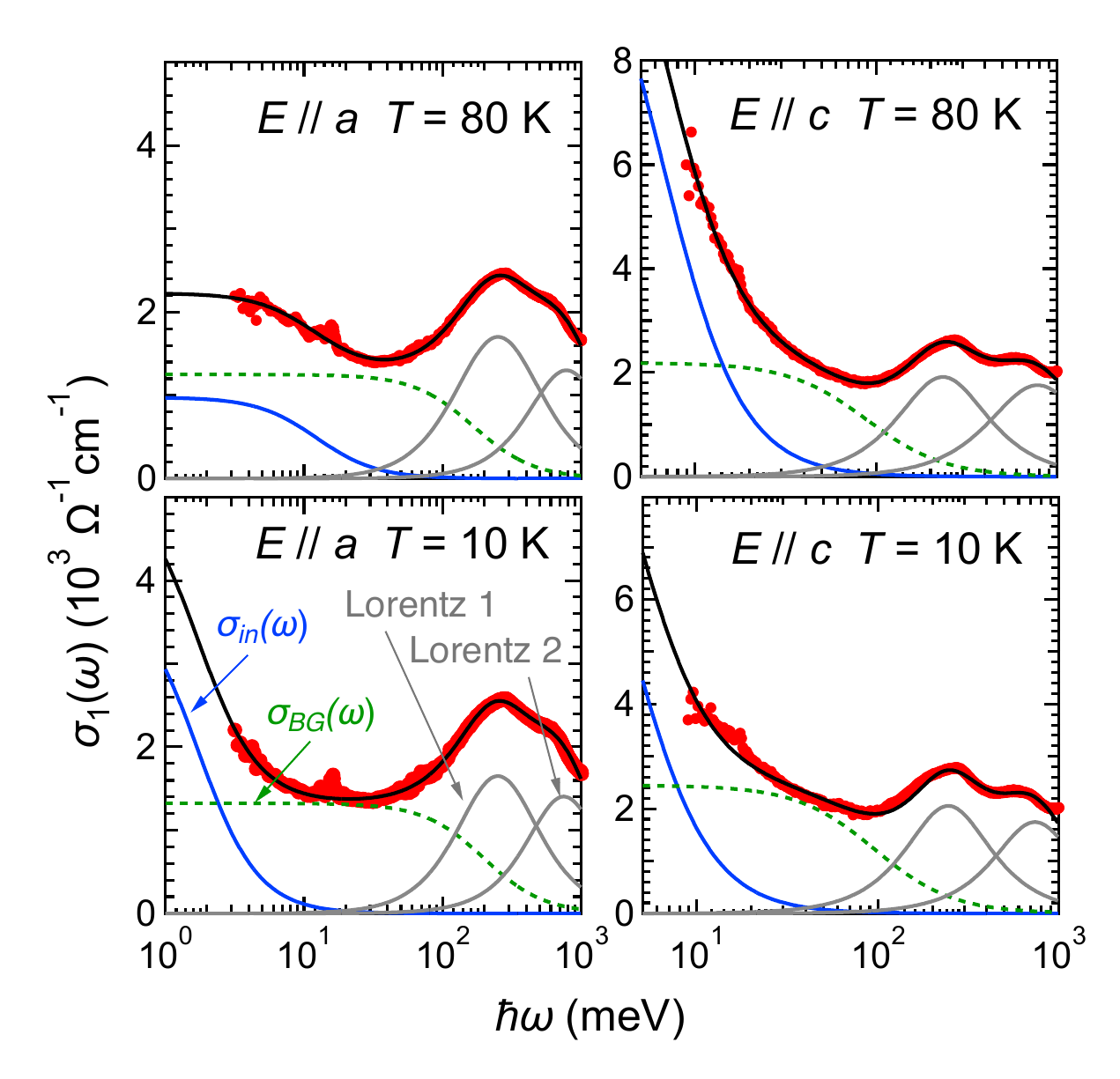}
\end{center}
\caption{
Low-energy part of optical conductivity spectra [$\sigma_1(\omega)$] of CeRhSn at $10$ and $80$~K for $E \parallel a$ and $E \parallel c$, and the fitting results with two-Drude and two-Lorentzian functions.
Two Drude components consist of a heavy quasiparticle component [$\sigma_{in}(\omega)$] and a light quasiparticle component [$\sigma_{BG}(\omega)$].
Two Lorentzians are assumed as the interband transitions from the valence band to the \cf hybridization bands with the Ce~$4f$ spin-orbit pair.
The fitting parameters are shown in Table~\ref{tbl:DL}.
}
\label{fig:S2}
\end{figure}

\begin{table*}
\caption{Fitting parameters of the Drude--Lorentz model of Eq.~\ref{eq:DL} attributed to the experimental $\sigma_1(\omega)$ spectra in Fig.~S2.}
\begin{center}
\begin{tabular}{|c|c|c|c|c|c|} \hline
   \multicolumn{2}{|c|}{} & \multicolumn{2}{c|}{$E \parallel a$} & \multicolumn{2}{c|}{$E \parallel c$} \\ \hline
   \multicolumn{2}{|c|}{$T =$} & 10~K & 80~K & 10~K & 80~K \\ \hline
   \multirow{2}{*}{Heavy Drude $\tilde{\sigma}_{in}(\omega)$} & $N_{in}e^2/m_0$ [${\rm eV}/(\Omega \cdot {\rm cm})$] & 6.75 $\pm$ 0.06 & 9.68 $\pm$ 0.21 & 77.46 $\pm$ 0.30 & 79.59 $\pm$ 0.49 \\
                & $1/\tau_{in}$ [eV] & 1.72e-3 $\pm$ 0.01e-3 & 9.18e-3 $\pm$ 0.22e-3 & 9.16e-3 $\pm$ 0.15e-3 & 6.62e-3 $\pm$ 0.04e-3 \\ \hline
                
   \multirow{2}{*}{Light Drude $\tilde{\sigma}_{BG}(\omega)$} & $N_{BG}e^2/m_0$ [${\rm eV}/(\Omega \cdot {\rm cm})$] & 262.3 $\pm$ 91.3 & 233.2 $\pm$ 92.0 & 226.4 $\pm$ 4.5 & 194.6 $\pm$ 3.4 \\
                & $1/\tau_{BG}$ [eV] & 0.198 $\pm$ 0.069 & 0.181 $\pm$ 0.072 & 0.119 $\pm$ 0.003 & 0.0889 $\pm$ 0.0022 \\ \hline
                
   \multirow{3}{*}{Lorentz~1 $\tilde{\sigma}_{L,1}(\omega)$} & $N_{1}e^2/m_0$ [${\rm eV}/(\Omega \cdot {\rm cm})$] & 742.3 $\pm$ 136.0 & 725.4 $\pm$ 163.0 & 640.6 $\pm$ 17.5 & 664.7 $\pm$ 28.4 \\
                & $1/\tau_1$ [eV] & 0.450 $\pm$ 0.016 & 0.456 $\pm$ 0.024 & 0.340 $\pm$ 0.005 & 0.346 $\pm$ 0.009 \\
                &  $\omega_1$ [eV] & 0.249 $\pm$ 0.004 & 0.248 $\pm$ 0.006 & 0.245 $\pm$ 0.001 & 0.234 $\pm$ 0.002 \\ \hline
                
   \multirow{3}{*}{Lorentz~2 $\tilde{\sigma}_{L,2}(\omega)$} & $N_{2}e^2/m_0$ [${\rm eV}/(\Omega \cdot {\rm cm})$] & 1716.5 $\pm$ 82.2 & 1846.5 $\pm$ 130.0 & 1737.2 $\pm$ 42.6 & 2208.1 $\pm$ 93.5 \\
                & $1/\tau_2$ [eV] & 1.22 $\pm$ 0.023 & 1.354 $\pm$ 0.036 & 0.994 $\pm$ 0.019 & 1.254 $\pm$ 0.042 \\
                &  $\omega_2$ [eV] & 0.744 $\pm$ 0.011 & 0.763 $\pm$ 0.020 & 0.737 $\pm$ 0.002 & 0.780 $\pm$ 0.006 \\ \hline
                
\end{tabular}
\end{center}
\label{tbl:DL}
\end{table*}%

Along both axes, \OCbg and two Lorentzians do not change so much at different temperatures, but only \OCin component strongly changes.
Therefore, the experimental \OCin components, in which the sum of the components of \OCbg and two Lorentzians is subtracted from the experimental \OC spectra, were used for the dynamical Planckian scaling plot in Fig.~4.

\vspace{5mm}
\section*{S3.~DPS plot for all temperatures for $E \parallel a$}

\begin{figure}
\begin{center}
\includegraphics[width=0.4\textwidth]{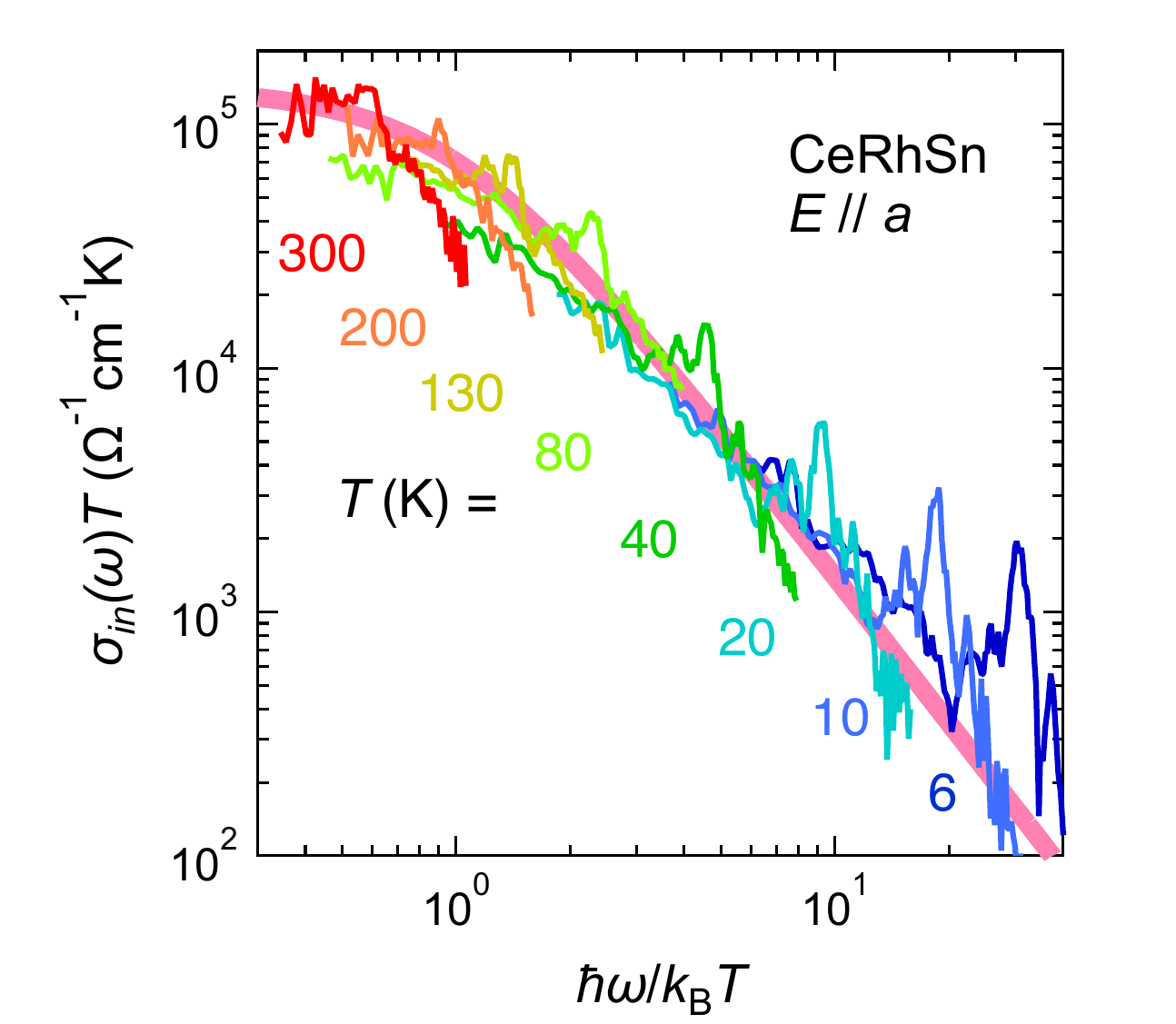}
\end{center}
\caption{
Same as Fig.~4(a), but the spectra for \EparaA at all temperatures are plotted.
The thick line is the theoretical DPS curve of $\sigma_1(\omega) \cdot T \propto [(\hbar \omega/k_{\rm B}T)^2+1]^{-1}$.
}
\label{fig:S2}
\end{figure}


Figure~S3 is the dynamical Planckian scaling (DPS) plot for all temperatures for \EparaA.
At all temperatures, the $\sigma_{in}(\omega) \cdot T$ lines are scaled in one universal line with a slight saturation at $\hbar \omega/ k_{\rm B} T \sim 1$.
Assuming a Drude formula, $\sigma_1(\omega) \propto \tau/[(\hbar \omega/k_{\rm B}T)^2+1]$, and the non-Fermi liquid character, $\tau \propto T^{-1}$, $\sigma_1(\omega) \cdot T \propto [(\hbar \omega/k_{\rm B}T)^2+1]^{-1}$.
The expected line plotted as a thick line in the figure can reproduce the experimental curves.
This result suggests that the DPS can follow at high temperatures.


%
%

\bibliographystyle{apsrev4-1}
\bibliography{CeRhSn_Notes}